
\input amssym  

\newdimen\normalparindent

\newdimen\paperwidth
\newdimen\paperheight
\paperwidth=210truemm
\paperheight=297truemm

\iffalse  

\hsize=15truecm
\hoffset=.46truecm
\vsize=23.7truecm
\voffset=.46truecm

\normalparindent=24pt

\font\elevenrm=cmr10 at 11pt 
\font\eightrm=cmr8 
\font\sixrm=cmr6 

\font\eleveni=cmmi10 at 11pt 
\font\eighti=cmmi8
\font\sixi=cmmi6

\font\elevensy=cmsy10 at 11pt 
\font\eightsy=cmsy8
\font\sixsy=cmsy6

\font\elevenex=cmex10 at 11pt 

\font\elevenbf=cmbx10 at 11pt 
\font\eightbf=cmbx8
\font\sixbf=cmbx6

\font\eleventt=cmtt10 at 11pt 
\font\elevensl=cmsl10 at 11pt 
\font\elevenit=cmti10 at 11pt 

\textfont0=\elevenrm \scriptfont0=\eightrm \scriptscriptfont0=\sixrm
\def\rm{\fam0\elevenrm}
\textfont1=\eleveni \scriptfont1=\eighti \scriptscriptfont1=\sixi
 
\textfont2=\elevensy \scriptfont2=\eightsy \scriptscriptfont2=\sixsy
\def\cal{\fam2}
\textfont3=\elevenex \scriptfont3=\elevenex \scriptscriptfont3=\elevenex
\textfont\itfam=\elevenit
\def\it{\fam\itfam\elevenit}
\textfont\slfam=\elevensl
\def\sl{\fam\slfam\elevensl}
\textfont\bffam=\elevenbf \scriptfont\bffam=\eightbf
\scriptscriptfont\bffam=\sixbf
\def\bf{\fam\bffam\elevenbf}
\textfont\ttfam=\eleventt
\def\tt{\fam\ttfam\eleventt}

\skewchar\eleveni='177 \skewchar\eighti='177 \skewchar\sixi='177
\skewchar\elevensy='60 \skewchar\eightsy='60 \skewchar\sixsy='60

\font\sc=cmcsc10 at 11pt 
\font\titlefont=cmssbx10 scaled \magstep3


\smallskipamount=3.5pt plus 1pt minus 1pt
\medskipamount=7pt plus 2pt minus 2pt
\bigskipamount=14pt plus 2pt minus 2pt
\normalbaselineskip=14pt
\normallineskip=1pt
\normallineskiplimit=0pt
\jot=3.5pt

\normalbaselines
\rm

\else  

\hsize=15truecm
\hoffset=.46truecm
\vsize=23.7truecm
\voffset=.46truecm

\normalparindent=20pt

\font\sc=cmcsc10


\font\titlefont=cmssbx10 scaled \magstep2

\fi

\def\S{\mathhexbox278\thinspace}

\def\square{\hbox to.77778em{%
\hfil\vrule\vbox to.675em{\hrule width.6em\vfil\hrule}\vrule\hfil}}

\long\def\acknowledgements#1\par{\medbreak\noindent{\it Acknowledgements\/}.\enspace#1\par\medbreak}
\def\definition#1\par{\medbreak\noindent{\bf Definition.}\enspace
  #1\par\medbreak}
\def\example#1\par{\medbreak\noindent{\bf Example.}\enspace
  #1\par\medbreak}
\def\question#1\par{\medbreak\noindent{\it Question.}\enspace
  #1\par\medbreak}
\long\def\remark#1\par{\medbreak\noindent{\it Remark\/}.\enspace#1\par\medbreak}
\def\exercise#1\par{\medbreak\noindent{\bf Exercise.}\enspace
#1\par\medbreak}
\def\notation#1\par{\medbreak\noindent{\bf Notation.}\enspace
#1\par\medbreak}
\def\proof{\noindent{\it Proof\/}.\enspace}
\def\endproof{\nobreak\hfill\quad\square\par\medbreak}

\def\lineover#1{{\offinterlineskip\mathchoice
{\setbox0=\hbox{$\displaystyle#1$}%
\vbox{\kern .33pt\hbox to\wd0{\kern 1pt\leaders\hrule height .33pt%
\hfill\kern 1pt}\kern 1pt\box0}}
{\setbox0=\hbox{$\textstyle#1$}%
\vbox{\kern .33pt\hbox to\wd0{\kern 1pt\leaders\hrule height .33pt%
\hfill\kern 1pt}\kern 1pt\box0}}
{\setbox0=\hbox{$\scriptstyle#1$}%
\vbox{\kern .25pt\hbox to\wd0{\kern .8pt\leaders\hrule height .25pt%
\hfill\kern .8pt}\kern .8pt\box0}}
{\setbox0=\hbox{$\scriptscriptstyle#1$}%
\vbox{\kern .2pt\hbox to\wd0{\kern .6pt\leaders\hrule height .2pt%
\hfill\kern .6pt}\kern .6pt\box0}}}}

\def\isom{\buildrel\sim\over\longrightarrow}

\def\isomorphism#1{\mathrel{\mathop{\longrightarrow}%
\limits^{#1}_{\raise0.5ex\hbox{$\scriptstyle\sim$}}}}

\def\injlim{\mathop{\vtop{\offinterlineskip\halign{##\cr
 \hfil\rm lim\hfil\cr\noalign{\kern.1ex}\rightarrowfill\cr
 \noalign{\kern-.4ex}\cr}}}}
\def\projlim{\mathop{\vtop{\offinterlineskip\halign{##\cr
 \hfil\rm lim\hfil\cr\noalign{\kern.1ex}\leftarrowfill\cr
 \noalign{\kern-.4ex}\cr}}}}

\def\smallmatrix#1#2#3#4{\bigl({#1\atop#3}\,{#2\atop#4}\bigr)}

\def\textfrac#1/#2{{\textstyle{#1\over#2}}}

\def\relativediag#1#2#3#4#5#6{\vcenter{\baselineskip=3ex \halign{
\hfil$##$&$##$&$##$\hfil\cr
#1\quad& \hfilneg\buildrel#2\over\longrightarrow\hfilneg& \quad#3\cr
\lower1ex\llap{$\scriptstyle#4\hskip-1ex$}\searrow& \quad&
\swarrow\lower1ex\rlap{$\hskip-1ex\scriptstyle#5$} \cr
& \hfilneg#6\hfilneg&\cr}}}

\def\trianglediag#1#2#3#4#5#6{\vcenter{\baselineskip=3ex \halign{
\hfil$##$&$##$&$##$\hfil\cr
#1\quad& \hfilneg\buildrel#2\over\longrightarrow\hfilneg& \quad#3\cr
\lower1ex\llap{$\scriptstyle#6\hskip-1ex$}\nwarrow& \quad&
\swarrow\lower1ex\rlap{$\hskip-1ex\scriptstyle#4$} \cr
& \hfilneg#5\hfilneg&\cr}}}

\def\correspondence#1#2#3#4#5{\vcenter{\baselineskip=3ex \halign{
\hfil$##$&$##$&$##$\hfil\cr
&\hfilneg#1\hfilneg\cr
\raise1ex\llap{$\scriptstyle#2$}\swarrow&&\searrow
\raise1ex\rlap{$\scriptstyle#3$}\cr
#4&&#5\cr}}}

\newif\iffirstpar
\everypar{\iffirstpar\parindent=\normalparindent\firstparfalse\fi}

\def\sectionheading#1{\subcount=0 \subsectioncount=0 \eqcount=0
  \bigskip\vskip\parskip
  \leftline{\bf #1}\nobreak\smallskip\firstpartrue\parindent=0pt}

\def\section#1\par{\advance\sectioncount by1%
  \edef\currentlabel{\number\sectioncount}%
  \sectionheading{\number\sectioncount.\enspace#1}}

\def\unnumberedsection#1\par{\sectionheading{#1}}

\def\subsection#1\par{\medbreak\penalty-200\advance\subsectioncount by1%
  \edef\currentlabel{\number\sectioncount.\number\subsectioncount}%
  \leftline{\it\number\sectioncount.\number\subsectioncount.\enspace#1}%
  \smallskip\parindent=0pt\firstpartrue}

\newwrite\auxfile

\newcount\sectioncount \sectioncount=0
\newcount\subsectioncount 
\newcount\subcount 
\newcount\eqcount 

\def\subno{\global\advance\subcount by1\relax
  \number\sectioncount.\number\subcount
  \xdef\currentlabel{\number\sectioncount.\number\subcount}}
\def\proclaim #1. #2\par{\medbreak
  \noindent{\bf#1~\subno.\enspace}{\sl#2\par}%
  \ifdim\lastskip<\medskipamount \removelastskip\penalty55\medskip\fi}
\def\proclaimx #1 (#2). #3\par{\medbreak
  \noindent{\bf#1~\subno\ \rm (#2).\enspace}{\sl#3\par}%
  \ifdim\lastskip<\medskipamount \removelastskip\penalty55\medskip\fi}

\newdimen\algindent
\def\plusindent{\advance\algindent by \parindent}
\def\minusindent{\advance\algindent by-\parindent}


\newcount\algstepcount

\long\def\algorithm (#1). #2\endalgorithm{\medbreak
  \algindent=0pt%
  \algstepcount=0%
  \noindent{\bf Algorithm~\subno} (#1). {\sl#2}\par\medbreak}

\def\step{\advance\algstepcount by1
\edef\currentlabel{\number\algstepcount}
\smallskip\hangindent\parindent
\advance\hangindent by\algindent\indent
\llap{{\bf \the\algstepcount.}\enspace}\kern\algindent
\ignorespaces}


\def\labeldef#1#2{\expandafter\gdef\csname L@#1\endcsname{#2}}
\def\label#1{%
  \expandafter\xdef\csname L@#1\endcsname{\currentlabel}%
  \write\auxfile{\string\labeldef{#1}{\csname L@#1\endcsname}}%
  \ignorespaces}
\def\ref#1{\expandafter\ifx\csname L@#1\endcsname\relax
  \message{Undefined label `#1'}??\else
  \csname L@#1\endcsname\fi}

\def\eqnumber#1{\global\advance\eqcount by1\relax
  \eqno(\number\sectioncount.\number\eqcount)%
  \expandafter\xdef\csname E@#1\endcsname{%
    \number\sectioncount.\number\eqcount}}
\def\eqref#1{\expandafter\ifx\csname E@#1\endcsname\relax
  \message{Undefined equation `#1'}??\else
  (\csname E@#1\endcsname)\fi}

\newcount\refcount \refcount=0
\def\citedef#1#2{\expandafter\gdef\csname C@#1\endcsname{#2}}
\def\cite#1{\expandafter\ifx\csname C@#1\endcsname\relax
  \message{Undefined reference `#1'}\citedef{#1}{??}\fi
  \expandafter\gdef\csname R@#1\endcsname{\relax}%
  [\csname C@#1\endcsname]}
\def\citex#1#2{\expandafter\ifx\csname C@#1\endcsname\relax
  \message{Undefined reference `#1'}\citedef{#1}{??}\fi
  \expandafter\gdef\csname R@#1\endcsname{\relax}%
  [\csname C@#1\endcsname, #2]}
\def\reference#1{\advance\refcount by 1%
  \expandafter\ifx\csname R@#1\endcsname\relax
  \message{Warning: reference `#1' not used}\fi
  \expandafter\edef\csname C@#1\endcsname{\the\refcount}%
  \write\auxfile{\string\citedef{#1}{\csname C@#1\endcsname}}%
  \item{[\csname C@#1\endcsname]}}

\newif\ifauxexists
\immediate\openin0=\jobname.aux
\ifeof 0
  \auxexistsfalse
\else
  \auxexiststrue
\fi
\immediate\closein0
\ifauxexists
  \input \jobname.aux
\else
  \message{No file `\jobname.aux'}
\fi
\openout\auxfile=\jobname.aux

\hyphenation{Klooster-man}

\def\C{{\bf C}}
\def\F{{\bf F}}
\def\cF{{\cal F}}
\def\Kl{\mathop{\rm Kl}\nolimits}
\def\M{{\cal M}}
\def\SU{\mathop{\rm SU}\nolimits}
\def\U{{\rm U}}
\def\Z{{\bf Z}}

\def\one{{\bf 1}}

\centerline{\titlefont On quantum computation of Kloosterman sums}

\bigskip

\centerline{Peter Bruin}
\smallskip
\centerline{3 October 2018}

\bigskip
{\narrower\narrower\noindent {\it Abstract.\/} We give two quantum
algorithms for computing (twisted) Kloosterman sums attached to a
finite field $\F$ of $q$ elements.  The first algorithm computes a
quantum state containing, as its coefficients with respect to the
standard basis, all Kloosterman sums for $\F$ twisted by a given
multiplicative character, and runs in time polynomial in $\log q$.
The second algorithm computes a single Kloosterman sum to a prescribed
precision, and runs in time quasi-linear in $\sqrt{q}$.\par}

\section Introduction

Let $p$ a prime number, and let $r$ a positive integer.  Let $\F$ be a
finite field of $q$ elements, where $q=p^r$.  An {\it additive
character\/} of~$\F$ is a group homomorphism
$$
\psi\colon \F\to\C^\times,
$$
and a {\it multiplicative character\/} of~$\F$ is a group homomorphism
$$
\chi\colon\F^\times\to\C^\times.
$$
There are many interesting number-theoretical constructions involving
such characters.

\definition Let $\psi$ be a non-trivial additive character of~$\F$.
For all $a\in \F^\times$ and all multiplicative characters $\chi$
of~$\F$, the {\it (twisted) Kloosterman sum\/} $\Kl_\psi(a,\chi)$ is
defined as
$$
\Kl_\psi(a,\chi)=\sum_{\textstyle{x,y\in \F\atop xy=a}}\chi(x)\psi(x+y).
$$

If $\chi$ is the trivial character, then $\Kl_\psi(a,\chi)$ is real.
In general, a straightforward computation shows that
$\Kl_\psi(a,\chi)$ satisfies
$$
\Kl_\psi(a,\chi) = \chi(-a)\overline{\Kl_\psi(a,\chi)},
\eqnumber{eq:conjugate}
$$
where the bar on the right-hand side denotes complex conjugation.  The
following (optimal) bound on absolute values of Kloosterman sums is a
famous result of Weil \cite{Weil}:
$$
|{\Kl_\psi(a,\chi)}|\le 2\sqrt{q}.
\eqnumber{Weil}
$$
No general closed formula for Kloosterman sums is known, except when
$q$ is odd and $\chi$ is the unique character
$\chi_2\colon\F^\times\to\{\pm1\}$ of order~$2$.  The Kloosterman sums
$\Kl_\psi(a,\chi_2)$ are also known as {\it Sali\'e sums\/} in honour
of the explicit formula
$$
\Kl_\psi(a,\chi_2)=G_\psi(\chi_2)
\sum_{\textstyle{y\in\F^\times\atop y^2=4a}} \psi(y)
$$
found by Sali\'e \cite{Salie}; here
$G_\psi(\chi_2)=\sum_{a\in\F^\times}\chi_2(a)\psi(a)$ is the Gauss sum
of the character $\chi_2$.

The problem of efficiently computing or approximating Kloosterman sums
does not seem to have a satisfactory solution as of this writing;
see \citex{Shparlinski}{Problem~3.45}
(= \citex{Shparlinski-web}{Problem~54}) and \citex{Childs}{\S8.1}.  In
this article, we are especially interested in {\it quantum\/}
algorithms.  An efficient (classical or quantum) algorithm for
approximating Kloosterman sums would have applications to quantum
algorithms for finding certain {\it hidden non-linear structures\/} as
in work of Childs, Schulman and
Vazirani \cite{Childs-Schulman-Vazirani}.

Although we have not been able to completely solve the problem of
efficiently computing Kloosterman sums, we do have the following two
results.  The first of these might be particularly applicable as a
building block in future quantum algorithms.

\proclaim Theorem. There exists a quantum algorithm that, given a
finite field\/~$\F$, a non-trivial additive character $\psi$ of\/~$\F$
and a multiplicative character $\chi$ of\/~$\F$, computes an $n$-qubit
state whose coefficients on the standard basis are the values
$\Kl_\psi(a,\chi)/\sqrt{N_\chi}$ for $a\in\F^\times$, and whose
running time is polynomial in $\log q$ as $q\to\infty$.  Here
$n=\lceil\log_2(q-1)\rceil$, and $N_\chi$ is an explicit positive
integer only depending on $\#\F$ and on whether the character~$\chi$
is trivial.

\label{theorem:vector}

In our second result, we bound the quantum complexity of computing the
normalised Kloosterman sums $\Kl_\psi(a,\chi)/\sqrt{q}$ to within a
prescribed absolute error.

\proclaim Theorem. There exists a quantum algorithm that, given a
finite field\/~$\F$, a non-trivial additive character $\psi$
of\/~$\F$, an element $a\in\F^\times$, a multiplicative character
$\chi$ of\/~$\F$ and real numbers $\delta,\epsilon\in(0,1)$, computes
an approximation to $\Kl_\psi(a,\chi)/\sqrt{q}$ that has absolute
error at most $\epsilon$ with probability at least $1-\delta$, and
whose running time is linear in $\sqrt{q}/(\delta\epsilon)$ times a
power of\/ $\log q$.

\label{theorem:single}

The main idea of the proof of Theorem~\ref{theorem:vector} is to
relate Kloosterman sums to Gauss sums.  We recall that the Gauss sum
of a multiplicative character~$\chi$ of~$\F$ (with respect to a fixed
non-trivial additive character~$\psi$) is defined as
$$
G_\psi(\chi)=\sum_{a\in\F^\times}\chi(a)\psi(a).
$$
Quantum computation of Gauss sums was studied by van Dam and
Seroussi \cite{van Dam-Seroussi}, who constructed a unitary operator
having Gauss sums (normalised to have absolute value 1) as
eigenvalues.  In fact, one of our ingredients is the main algorithm
of~\cite{van Dam-Seroussi}.

Our strategy, inspired by Katz \citex{Katz}{\S4.0}, is to use the {\it
multiplicative\/} Fourier transform (as opposed to the additive
Fourier transform used in \cite{van Dam-Seroussi}) to reduce the
problem of computing the function $a\mapsto\Kl_\psi(a,\chi)$ for a
given multiplicative character $\chi$ of~$\F$ to the problem of
computing the function mapping a multiplicative character $\chi'$
of~$\F$ to the complex number $G_\psi(\chi')G_\psi(\chi\chi')$.  In
order to deduce Theorem~\ref{theorem:single} from
Theorem~\ref{theorem:vector}, we use amplitude
amplification \cite{BHMT}.

\remark The problem of computing Kloosterman sums is a ``natural
problem'' in the sense that it does not involve a black box.  Other
natural problems for which efficient quantum algorithms have been
developed are Shor's algorithms for integer factorisation and discrete
logarithms \cite{Shor}, van Dam and Seroussi's algorithm for Gauss
sums mentioned above, and algorithms for computing unit groups and
ideal class groups of number fields; see \cite{Hallgren}, \cite{EHKS}
and \cite{Biasse-Song}.

To put the above theorems in a broader perspective, we briefly discuss
equidistribution properties of Kloosterman sums.  Although these
results are not used in the remainder of this article, they could be
used as a statistical check on possible future implementations.

We first consider the case where $\chi\colon\F^\times\to\C^\times$ is
the trivial character~$\one$, so we consider the Kloosterman sums
$$
\Kl_\psi(a,\one)
= \sum_{\textstyle{x,y\in\F\atop xy=a}}\psi(x+y).
$$

\proclaimx Theorem (Katz \citex{Katz}{Chapter~13}). For each prime
power $q$, let\/ $\F_q$ be a finite field of $q$ elements, and let
$\psi_q$ be a non-trivial additive character of\/~$\F_q$.  As
$q\to\infty$, the normalised Kloosterman sums
$\Kl_{\psi_q}(a,\one)/\sqrt{q}$ for $a\in\F^\times$ become
equidistributed with respect to the probability measure ${1\over
2\pi}\sqrt{4-x^2}\,dx$ on $[-2,2]$.

\label{theorem:equidistribution-real}

The measure in Theorem~\ref{theorem:equidistribution-real} is the
Sato--Tate measure for the compact Lie group $\SU(2)$; this is the
distribution of traces of elements of $\SU(2)$ that are randomly
distributed with respect to the Haar measure.  In view of Weil's
bound \eqref{Weil} and the fact that $\Kl_\psi(a,\one)$ is real, we
can write
$$
\Kl_\psi(a,\one) = 2\sqrt{q}\cos(\theta_\psi(a))
$$
for a unique angle $\theta_\psi(a)\in[0,\pi]$.
Theorem~\ref{theorem:equidistribution-real} can equivalently be stated
by saying that as $q\to\infty$, the angles $\theta_\psi(a)$ for
$a\in\F^\times$ become equidistributed with respect to the probability
measure ${2\over\pi}(\sin\theta)^2d\theta$ on $[0,\pi]$.

A similar equidistribution result holds when the character~$\chi$ is
allowed to vary; this was kindly explained to the author by Nicholas
M. Katz.

\proclaimx Theorem (Katz, personal communication). For each prime
power $q$, let\/ $\F_q$ be a finite field of $q$ elements, and let
$\psi_q$ be a non-trivial additive character of\/~$\F_q$.  As
$q\to\infty$, the normalised Kloosterman sums
$\Kl_{\psi_q}(a,\chi)/\sqrt{q}$, for $a\in\F_q^\times$ and $\chi$ a
multiplicative character of\/~$\F_q$, become equidistributed with
respect to the probability measure on the complex disc of radius~$2$
given in polar coordinates $(r,\theta)$ by ${1\over
2\pi^2}\sqrt{4-r^2}\,dr\,d\theta$.

\label{theorem:equidistribution}

The measure in Theorem~\ref{theorem:equidistribution} is the
Sato--Tate measure for the compact Lie group $\U(2)$.  It follows
from \eqref{eq:conjugate} and \eqref{Weil} that the two roots of the
polynomial
$$
f_\psi(a,\chi) = t^2 - {\Kl_\psi(a,\chi)\over\sqrt{q}} t + \chi(-a) \in \C[t]
$$
have absolute value 1, so there are two angles $\theta_\psi(a,\chi)$,
$\theta'_\psi(a,\chi)$ (unique up to ordering) satisfying
$$
\eqalign{
\Kl_\psi(a,\chi) &= \sqrt{q}\bigl(\exp(i\theta_\psi(a,\chi))
+\exp(i\theta'_\psi(a,\chi))\bigr),\cr
\chi(-a) &= \exp\bigl(i(\theta_\psi(a,\chi)+\theta'_\psi(a,\chi))\bigr).}
$$
(For $\chi=\one$, these angles are $\pm\theta_\psi(a)$.)  In other
words, there is a unique conjugacy class in $\U(2)$ having
characteristic polynomial $f_\psi(a,\chi)$.  Katz's proof of
Theorem~\ref{theorem:equidistribution} shows that as $q\to\infty$,
these conjugacy classes are equidistributed for the probability
measure on the space of conjugacy classes induced by the Haar measure
on $\U(2)$.

We end this introduction with two questions that appear natural to ask
in the light of the above results.

\question Can one find a family of curves or higher-dimensional
varieties over finite fields for which knowing the quantum state from
Theorem~\ref{theorem:vector} leads to an efficient algorithm for
counting points?

\question Can one construct a unitary operator that has (a subset of)
the numbers $\exp(i\theta_\psi(a,\chi))$ and
$\exp(i\theta'_\psi(a,\chi))$ as eigenvalues and that can be
efficiently implemented on a quantum computer?  The construction of
such an operator will necessarily be closely connected to a proof of
Weil's bound and to making the Kloosterman sheaves of Katz~\cite{Katz}
computationally accessible.

\acknowledgements I would like to thank Serge Fehr for his comments on
a draft version of this article.  I am also grateful to Javier
Fres\'an, Richard Griffon and Igor Shparlinski for useful discussions
about Kloosterman sums.  Finally, it is a pleasure to thank Nicholas
M. Katz for providing a proof of
Theorem~\ref{theorem:equidistribution}.

\section Preliminaries

Let $X_\F$ denote the group of multiplicative characters of~$\F$,
i.e. group homomorphisms $\chi\colon \F^\times\to\C^\times$.  For any
function $f\colon \F^\times\to\C$, the {\it multiplicative Fourier
transform\/} of~$f$ is the function
$$
\eqalign{
\M f\colon X_\F&\longrightarrow\C\cr
\chi&\longmapsto\sum_{a\in \F^\times}\chi(a)f(a).}
$$
This defines a map
$$
\M\colon \C^{\F^\times}\longrightarrow\C^{X_\F}.
$$
This is an isomorphism of $\C$-vector spaces; the inverse $\M^{-1}$
maps a function $\phi\colon X_\F\to\C$ to the function
$$
\eqalign{
\M^{-1}\phi\colon \F^\times&\longrightarrow\C\cr
a&\longmapsto{1\over q-1}\sum_{\chi\in X_\F}\chi(a)^{-1}\phi(\chi).}
$$

In particular, for all functions $f\colon \F^{\times}\to\C$ we have
the identity
$$
f = \M^{-1}\M f = {1\over q-1}\sum_{\chi\in X_\F}\M f(\chi)\chi^{-1}
\quad\hbox{in }\C^{\F^\times},
$$
and for all functions $\phi\colon X_\F\to\C$ we have the identity
$$
\phi = \M\M^{-1}\phi = \sum_{a\in \F^\times}\M^{-1}\phi(a) \hat a
\quad\hbox{in }\C^{X_\F},
$$
where $\hat a\in\C^{X_\F}$ is the function $X_\F\to\C$ sending $\chi$
to $\chi(a)$.

For two functions $f,g\colon \F^\times\to\C$, the {\it convolution\/}
of $f$ and~$g$ is
$$
\eqalign{
f*g\colon \F^\times &\longrightarrow \C\cr
a&\longmapsto\sum_{\textstyle{x,y\in \F^\times\atop xy=a}}f(x)g(y).}
$$
An elementary computation shows that
$$
\M f(\chi)\M g(\chi)=\M(f*g)(\chi)
\quad\hbox{for all }\chi\in X_\F.
$$

If $\psi$ is a non-trivial additive character of~$\F$ and $\chi$ is a
(possibly trivial) multiplicative character of~$\F$, then the
definition of the Gauss sum $G_\psi(\chi)$ immediately implies
$$
\M\psi(\chi) = G_\psi(\chi).
\eqnumber{M=G}.
$$

\section Encodings into quantum states

\def\ket#1{\mathopen|#1\mathclose\rangle}

For all $n\ge0$, let $V_n$ denote the $2^n$-dimensional Hilbert space
of $n$-qubit states; this can be identified with $(\C^2)^{\otimes n}$
equipped with the standard Hermitean metric.  The standard basis
vectors are labelled $\ket{0}$, $\ket{1}$, \dots, $\ket{2^n-1}$.  When
we are dealing with states inside a tensor product of the form
$V_{n_1}\otimes\cdots\otimes V_{n_k}$, we will refer to the $i$-th
tensor factor as the $i$-th {\it register\/}.

The quantum Fourier transform modulo~$q-1$ is the unitary
transformation $\cF_{q-1}$ on $n$-qubit states defined on the subspace
$\C\ket{0}+\cdots+\C\ket{q-2}$ by
$$
\cF_{q-1}\ket m =
{1\over\sqrt{q-1}}\sum_{d=0}^{q-2}\exp(2\pi i dm/(q-1))\ket d
$$
and extended in an unspecified way to the orthogonal complement.
This can be computed exactly using an efficient quantum
algorithm \cite{Mosca-Zalka}, and we will use this for simplicity.  In
practice, an approximate Fourier transform would probably be more
useful.

We fix generators $a_1$ and $\chi_1$ of the finite cyclic groups
$\F^\times$ and $X_\F$, respectively, that are assumed to satisfy
$$
\chi_1(a_1)=\exp(2\pi i/(q-1)).
$$
This choice determines isomorphisms
$$
\eqalign{
\Z/(q-1)\Z&\isom \F^\times\cr
d&\longmapsto a_1^d}
$$
and
$$
\eqalign{
\Z/(q-1)\Z&\isom X_\F\cr
m&\longmapsto\chi_1^m}
$$

\goodbreak

For all $a\in \F^\times$, we write $d_a$ for the discrete logarithm
of~$a$ with respect to~$a_1$, i.e.\ the unique
$d\in\{0,1,\ldots,q-2\}$ such that $a_1^d=a$.  Similarly, for all
$\chi\in X_\F$, we write $m_\chi$ for the unique $m\in\{0,1,\ldots,
q-2\}$ such that $\chi_1^m=\chi$.  Then for all $a\in \F^\times$ and
$\chi\in X_\F$, we have
$$
\chi(a)=\exp(2\pi i m_\chi d_a/(q-1)).
$$

Let $n$ be the least integer such that $2^n\ge q-1$.  For all
$a\in\F^\times$, we write $\ket{a}_\times$ for the $n$-qubit state
$\ket{d_a}$, and for all $\chi\in X_\F$ we write $\ket{\chi}_*$ for
the $n$-qubit state $\ket{m_\chi}$.  Furthermore, let $l$ be the least
integer such that $2^l\ge p$.  We assume that we are given a basis
$(b_0,\ldots,b_{r-1})$ for $\F$ as an $\F_p$-vector space.  We encode
elements of~$\F_p$ as strings of $l$ bits, and elements of~$\F$ by
their coefficients with respect to $(b_0,\ldots,b_{r-1})$.  In this
way we get an encoding of elements of $\F$ as strings of $lr$ bits,
and hence as $lr$-qubit states.  We write $\ket{x}_+$ for the state
defined by $x\in\F$ in this way.  For $a\in\F^\times$ we will use both
representations $\ket{a}_+$ and $\ket{a}_\times$.

We further assume that $\psi$ is given to us as the list of values
$\psi(b_0)$, \dots, $\psi(b_{r-1})$.  Then we can efficiently compute
the unitary operator
$$
U_\psi\colon V_{lr}\to V_{lr}
$$
such that
$$
U_\psi\ket{x}_+=\psi(x)\ket{x}_+\quad\hbox{for all }x\in \F.
\eqnumber{U-psi}
$$
We can also efficiently compute the Fourier transform
$$
\cF_{\F,\psi}\colon V_{lr}\to V_{lr}
$$
defined on states of the form $\ket{x}_+$ by
$$
\cF_{\F,\psi}\ket{x}_+ = {1\over\sqrt{q}}\sum_{y\in\F}\psi(xy)\ket{y}_+
$$
and extended in an unspecified way to the orthogonal complement.
Furthermore, we can efficiently compute the unitary operator
$$
W\colon V_n\otimes V_n\longrightarrow V_n\otimes V_n
$$
mapping any state $\ket{\chi}_*\ket{a}_\times=\ket{m}\ket{d}$ for
$\chi=\chi_1^d\in X_\F$ and $a=a_1^d\in\F^\times$ to the state
$\exp(2\pi md/(q-1))\ket{m}\ket{d}
= \chi(a)\ket{\chi}_*\ket{a}_\times$.

Using classical operations in the finite field~$\F$, we can
efficiently compute the unitary operator sending any state
$\ket{a}_\times\ket{x}_+=\ket{d}\ket{x}_+$ for $a=a_1^d\in\F^\times$
and $x\in\F$ to the state
$\ket{d}\ket{a_1^d x}_+=\ket{a}_\times\ket{ax}_+$.  Thanks to Shor's
algorithm for discrete logarithms \cite{Shor} and the exact version
from~\cite{Mosca-Zalka}, we furthermore have efficiently computable
operators
$$
\eqalign{
\exp_{a_1}\colon V_n\otimes V_{lr}&\longrightarrow V_n\otimes V_{lr},\cr
\log_{a_1}\colon V_n\otimes V_{lr}&\longrightarrow V_n\otimes V_{lr}}
$$
such that for $0\le d\le q-2$ we have
$$
\exp_{a_1}(\ket d\ket 0)=\ket 0\ket{a_1^d}_+
$$
and
$$
\log_{a_1}(\ket 0\ket{a_1^d}_+)=\ket d\ket 0.
$$
Using the convention that quantum registers that are initialised to
zero (or reset to zero by ``uncomputing'' a value computed earlier)
are omitted from the notation, we will abbreviate this as
$$
\exp_{a_1}(\ket{a}_\times)=\ket{a}_+
$$
and
$$
\log_{a_1}(\ket{a}_+)=\ket{a}_\times.
$$

\section Computing Gauss sums

Let $\one\in X_\F$ be the trivial multiplicative character of~$\F$.
It is well known that for all multiplicative characters $\chi\in
X_\F$, the Gauss sum $G_\psi(\chi)$ has absolute value
$$
|G_\psi(\chi)|=\sqrt{L_\chi},
$$
where
$$
L_\chi=\cases{
q& if $\chi\ne\one$,\cr
1& if $\chi=\one$.}
\eqnumber{G-abs}
$$
Furthermore, we have
$$
G_\psi(\one)=-1.
$$
In particular, \eqref{G-abs} implies
$$
\sum_{\chi\in X_\F}|G_\psi(\chi)|^2=(q-2)q+1=(q-1)^2.
\eqnumber{L2-G}
$$

We sketch a variant of the algorithm of van Dam and Seroussi \cite{van
Dam-Seroussi} that realises the Gauss sum $G_\psi(\chi)$ for a
non-trivial element $\chi\in X_\F$ as a phase shift.  This algorithm
is based on the fact that for all $y\in\F$ we have the identity
$$
\sum_{a\in\F^\times}\chi(a)\psi(ay)=\cases{
\chi(y)^{-1}G_\psi(\chi)& if $y\in\F^\times$,\cr
0& if $y=0$.}
\eqnumber{identity}
$$

\algorithm (Compute a single Gauss sum). Given a finite field\/~$\F$,
a non-trivial additive character $\psi$ of\/~$\F$ and a quantum state
$\ket{\chi}_*$ encoding a non-trivial multiplicative character $\chi$
of\/~$\F$, this algorithm outputs the state
$q^{-1/2}G_\psi(\chi)\ket{\chi}_*$.

\step Start with the state
$\ket{\chi}_*\ket{1}_\times=\ket{\chi}_*\ket{0}$.

\step Apply $\cF_{q-1}$ to the second register to obtain the state
${1\over\sqrt{q-1}}\sum_{a\in\F^\times}\ket{\chi}_*\ket{a}_\times$.

\step Apply $W$ to obtain the state
${1\over\sqrt{q-1}}\sum_{a\in\F^\times}\chi(a)\ket{\chi}_*\ket{a}_\times$.

\step Apply $\exp_{a_1}$ to the second register to obtain the state
${1\over\sqrt{q-1}}\sum_{a\in\F^\times}\chi(a)\ket{\chi}_*\ket{a}_+$.

\step Apply the Fourier transform $\cF_{\F,\psi}$ to obtain the state
$$
\eqalign{
{1\over\sqrt{q(q-1)}}\sum_{y\in\F}\biggl(
\sum_{a\in\F^\times}
\chi(a)\psi(ay)\biggr)\ket{\chi}_*\ket{y}_+
&={1\over\sqrt{q(q-1)}}G_\psi(\chi)
\sum_{y\in\F^\times}\chi(y)^{-1}\ket{\chi}_*\ket{y}_+,}
$$
where the equality follows from~\eqref{identity}.

\step Apply $\log_{a_1}$ to the second register to obtain the state
${1\over\sqrt{q(q-1)}}G_\psi(\chi)
\sum_{y\in\F^\times}\chi(y)^{-1}\ket{\chi}_*\ket{y}_\times$.

\step Apply $W$ to obtain the state
${1\over\sqrt{q(q-1)}}G_\psi(\chi)\sum_{y\in\F^\times}\ket{\chi}_*\ket{y}_\times$.

\step Apply $\cF_{q-1}^\dagger$ to the second register to obtain the
state $q^{-1/2}G_\psi(\chi)\ket{\chi}_*\ket{1}_\times$.

\endalgorithm

\label{algorithm:vDS}

In fact, it turns out to be slightly easier to give an algorithm that
computes all Gauss sums $G_\psi(\chi)$ for $\chi\in X_\F$
simultaneously, using the multiplicative rather than the additive
Fourier transform to make the Gauss sums appear.  In our algorithm for
computing Kloosterman sums, we will use both this algorithm and the
algorithm of van Dam and Seroussi.  We define the $n$-qubit state
$$
\ket{G_\psi}={1\over q-1}\sum_{\chi\in X_\F} G_\psi(\chi)\ket{\chi}_*;
\eqnumber{G-psi}
$$
note that this state is normalised because of~\eqref{L2-G}.

\algorithm (Compute the vector of Gauss sums). Given a finite
field\/~$\F$ and a non-trivial additive character $\psi$ of\/~$\F$,
this algorithm outputs the state $\ket{G_\psi}$.

\step Start with the state $\ket{1}_\times=\ket{0}$.

\step Apply $\cF_{q-1}$ to obtain the state
${1\over\sqrt{q-1}}\sum_{a\in \F^\times}\ket{a}_\times$.

\step Apply $\exp_{a_1}$ to obtain the state
${1\over\sqrt{q-1}}\sum_{a\in \F^\times}\ket{a}_+$.

\step Apply the operator $U_\psi$ defined in~\eqref{U-psi} to obtain
the state ${1\over\sqrt{q-1}}\sum_{a\in \F^\times}\psi(a)\ket{a}_+$.

\step Apply $\log_{a_1}$ to obtain the state
${1\over\sqrt{q-1}}\sum_{a\in \F^\times}\psi(a)\ket{a}_\times$.

\step Apply $\cF_{q-1}$ to obtain the state
${1\over q-1}\sum_{\chi\in X_\F}\M\psi(\chi)\ket{\chi}_* = {1\over
q-1}\sum_{\chi\in X_\F}G_\psi(\chi)\ket{\chi}_* = \ket{G_\psi}$, where
the equalities follow from \eqref{M=G} and~\eqref{G-psi}.

\endalgorithm

\label{algorithm:G-psi}

\section Computing Kloosterman sums

In this section, we prove Theorem~\ref{theorem:vector} by exhibiting
an algorithm for computing Kloosterman sums as in the theorem.  Our
algorithm is based on the following observation.

\proclaim Proposition. Let $\F$ be a finite field, let $\psi$ be a
non-trivial additive character of\/~$\F$, and let $\chi$ be a
multiplicative character of\/~$\F$.  The multiplicative Fourier
transform of the function $\F^\times\to\C$ mapping $a$ to
$\Kl_\psi(a,\chi)$ is the function
$$
\eqalign{
\Gamma_\psi^\chi\colon X_\F&\longrightarrow\C\cr
\chi'&\longmapsto
\sum_{a\in\F^\times}\Kl_\psi(a,\chi)\chi'(a)=
G_\psi(\chi\chi')G_\psi(\chi').}
$$

\label{prop:key}

\proof This is a special case of \citex{Katz}{Scholium
4.0.1}.\endproof

We write
$$
N_\chi = \sum_{a\in\F^\times}|{\Kl_\psi(a,\chi)}|^2.
$$
A straightforward computation using Proposition~\ref{prop:key},
Parseval's identity (i.e.\ the fact that the operator $(q-1)^{-1/2}\M$
is unitary) and \eqref{G-abs} yields
$$
N_\chi=\cases{
q^2-q-1& if $\chi=\one$,\cr
q^2-2q& if $\chi\ne\one$.}
$$
Note that the last case cannot occur if $q=2$, so $N_\chi$ is strictly
positive.

Given a non-trivial additive character $\psi$ of~$\F$ and a (possibly
trivial) multiplicative character $\chi$ of~$\F$, we define the
$n$-qubit state
$$
\ket{\Kl_\psi^\chi}={1\over\sqrt{N_\chi}}\sum_{a\in\F^\times}
\Kl_\psi(a,\chi)\ket a.
\eqnumber{eq:Kl}
$$

\goodbreak

We also define two $n$-qubit states
$$
\eqalign{
\ket{\Gamma_\psi^\chi} &=
{1\over\sqrt{(q-1)N_\chi}}\sum_{\chi'\in X_\F}
\Gamma_\psi^\chi(\chi')\ket{\chi'}_*,\cr
\ket{\widetilde\Gamma_\psi^\chi} &=
{1\over q-1}\biggl(
\Gamma_\psi^\chi(\one)\ket{\one}_*
+{1\over\sqrt{q}}\sum_{\chi'\in X_\F\setminus\{\one\}}
\Gamma_\psi^\chi(\chi')\ket{\chi'}_*\biggr).}
$$
By Proposition~\ref{prop:key}, we have
$$
\cF_{q-1}\ket{\Kl_\psi^\chi} = \ket{\Gamma_\psi^\chi},
$$
or equivalently
$$
\cF_{q-1}^\dagger \ket{\Gamma_\psi^\chi} =\ket{\Kl_\psi^\chi},
$$
which shows that computing $\ket{\Kl_\psi^{\chi}}$ is equivalent to
computing $\ket{\Gamma_\psi^\chi}$.  However, the state
$\ket{\widetilde\Gamma_\psi^\chi}$ appears to be easier to obtain
(starting from just $\psi$ and $\chi$) than $\ket{\Gamma_\psi^\chi}$;
see Algorithm~\ref{algorithm:intermediate} below.  That having been
said, the states $\ket{\Gamma_\psi^\chi}$ and
$\ket{\widetilde\Gamma_\psi^\chi}$ are very close for large $q$, as we
will now show.

\proclaim Lemma. The asymptotic behaviour of the inner product
$\langle{\widetilde\Gamma_\psi^\chi}\mid{\Gamma_\psi^\chi}\rangle$ as
$q\to\infty$ is given by
$$
\langle{\widetilde\Gamma_\psi^\chi}\mid{\Gamma_\psi^\chi}\rangle
=\cases{
1-{1\over2}q^{-2}+O(q^{-5/2})& if $\chi=\one$,\cr
1-{1\over2}q^{-1}+O(q^{-3/2})& if $\chi\ne\one$.}
$$

\proof First, we note that
$$
|\Gamma_\psi^\chi(\one)|^2=|G_\psi(\chi)G_\psi(\one)|^2=L_\chi,
$$
with $L_\chi$ as in~\eqref{G-abs}.  Furthermore, we have
$$
\sum_{\chi'\in X_\F\setminus\{\one\}}|\Gamma_\psi^\chi(\chi')|^2=q M_\chi,
$$
where
$$
M_\chi=\cases{
q^2-2q& if $\chi=\one$,\cr
q^2-3q+1& if $\chi\ne\one$.}
$$
In view of this, we define two $n$-qubit states
$$
\eqalign{
\ket{\Gamma_\psi^\chi}_+ &= {1\over\sqrt{L_\chi}}
\Gamma_\psi^\chi(\one)\ket{\one}_*,\cr
\ket{\Gamma_\psi^\chi}_- &= {1\over\sqrt{q M_\chi}}
\sum_{\chi'\in X_\F\setminus\{\one\}} \Gamma_\psi^\chi(\chi')\ket{\chi'}_*.}
$$
The state $\ket{\widetilde\Gamma_\psi^\chi}$ can be decomposed as
$$
\ket{\widetilde\Gamma_\psi^\chi} = \alpha^\chi_+\ket{\Gamma_\psi^\chi}_+
+ \alpha^\chi_-\ket{\Gamma_\psi^\chi}_-,
$$
where
$$
\alpha^\chi_+={\sqrt{L_\chi}\over q-1}
\quad\hbox{and}\quad
\alpha^\chi_-={\sqrt{M_\chi}\over q-1}.
\eqnumber{eq:alpha}
$$
Similary, the state $\ket{\Gamma_\psi^\chi}$ can be decomposed as
$$
\ket{\Gamma_\psi^\chi} = \beta^\chi_+\ket{\Gamma_\psi^\chi}_+
+ \beta^\chi_-\ket{\Gamma_\psi^\chi}_-,
$$
where
$$
\beta^\chi_+=\sqrt{L_\chi\over (q-1)N_\chi}
\quad\hbox{and}\quad
\beta^\chi_-=\sqrt{q M_\chi\over (q-1)N_\chi}.
\eqnumber{eq:beta}
$$
Expanding $\alpha^\chi_\pm$ and $\beta^\chi_\pm$ in power series in
$q^{-1/2}$ and substituting these in the equality
$$
\langle{\widetilde\Gamma_\psi^\chi}\mid{\Gamma_\psi^\chi}\rangle
=\alpha^\chi_+\beta^\chi_++\alpha^\chi_-\beta^\chi_-,
$$
we obtain the claim.\endproof

We will now give an algorithm for computing
$\ket{\widetilde\Gamma_\psi^\chi}$ and then show how one can use
amplitude amplification to transform this into an algorithm for
computing $\ket{\Gamma_\psi^\chi}$.

\algorithm (Compute the state $\ket{\widetilde\Gamma_\psi^\chi}$).
Given a finite field $\F$, a non-trivial additive character $\psi$
of\/~$\F$ and a quantum state $\ket{\chi}_*$ encoding a (possibly
trivial) multiplicative character $\chi$ of\/~$\F$, this algorithm
outputs the state $\ket{\chi}_*\ket{\widetilde\Gamma_\psi^\chi}$.

\step Apply Algorithm~\ref{algorithm:G-psi} to an auxiliary $n$-qubit
register to compute the state
$$
\ket{\chi}_*\ket{G_\psi} = {1\over q-1} \sum_{\chi'\in X_\F}
G_\psi(\chi')\ket{\chi}_*\ket{\chi'}_*.
$$

\step Apply the operator
$\ket{\chi}_*\ket{\chi'}_* \mapsto \ket{\chi}_*\ket{\chi^{-1}\chi'}_*$
to obtain the state
$$
{1\over q-1}\sum_{\chi'\in X_\F}G_\psi(\chi')\ket{\chi}_*\ket{\chi^{-1}\chi'}_*
={1\over q-1}\sum_{\chi'\in X_\F}G_\psi(\chi\chi')\ket{\chi}_*\ket{\chi'}_*.
$$

\step Apply either a phase change by $-1$ (for $\chi'=\one$) or
Algorithm~\ref{algorithm:vDS} (for $\chi'$ non-trivial) to the second
register to obtain the state
$\ket{\chi}_*\ket{\widetilde\Gamma_\psi^\chi}$.

\endalgorithm

\label{algorithm:intermediate}

For all $\chi\in X_\F$, let $Z_\psi^\chi$ denote the unitary operator
on the second $n$-qubit register defined by the above algorithm, so we
have in particular
$$
Z_\psi^\chi\ket{0}=\ket{\widetilde\Gamma_\psi^\chi}
=\alpha^\chi_+\ket{\Gamma_\psi^\chi}_+ + \alpha^\chi_-\ket{\Gamma_\psi^\chi}_-.
$$
We define three angles $\theta_\chi,\omega_\chi,\rho_\chi$ by
$$
\displaylines{
\sin\theta_\chi = \alpha^\chi_+,\quad\cos\theta_\chi = \alpha^\chi_-,\cr
\sin\omega_\chi = \beta^\chi_+,\quad\cos\omega_\chi = \beta^\chi_-,\cr
\rho_\chi=\theta_\chi-\omega_\chi.}
$$
Then we have
$$
\ket{\Gamma_\psi^\chi} =
\beta^\chi_+\ket{\Gamma_\psi^\chi}_+ + \beta^\chi_-\ket{\Gamma_\psi^\chi}_-
= R_\chi \ket{\widetilde\Gamma_\psi^\chi},
$$
where $R_\chi$ is the unitary operator on
$\C\ket{\Gamma_\psi^\chi}_++\C\ket{\Gamma_\psi^\chi}_-$ defined by the
matrix
$\smallmatrix{\cos\rho_\chi}{-\sin\rho_\chi}{\sin\rho_\chi}{\cos\rho_\chi}$
with respect to the basis
$(\ket{\Gamma_\psi^\chi}_+,\ket{\Gamma_\psi^\chi}_-)$.

The angle~$\rho_\chi$ satisfies the inequality
$$
|{\sin\rho_\chi}|\le \sin(2\theta_\chi)
$$
or equivalently
$$
|\alpha^\chi_+\beta^\chi_--\alpha^\chi_-\beta^\chi_+|
\le 2\alpha^\chi_+\alpha^\chi_-,
$$
which is not hard to verify using \eqref{eq:alpha}
and~\eqref{eq:beta}.
Using the techniques of H{\o}yer \citex{Hoyer}{\S III}, we therefore
see that $R_\chi$ can be implemented using one application of
$Z_\psi^\chi$, one application of $(Z_\psi^\chi)^\dagger$ and a number
of (conditional) phase shifts.

\algorithm (Compute the vector of Kloosterman sums). Given a finite
field $\F$, a non-trivial additive character $\psi$ of\/~$\F$ and a
quantum state $\ket{\chi}_*$ encoding a (possibly trivial)
multiplicative character $\chi$ of\/~$\F$, this algorithm outputs the
state $\ket{\chi}_*\ket{\Kl_\psi^\chi}$.

\step Apply Algorithm~\ref{algorithm:intermediate} to obtain the state
$\ket{\chi}_*\ket{\widetilde\Gamma_\psi^\chi}$.

\step Apply the operator $R_\chi$ to the second register to obtain the
state $\ket{\chi}_*\ket{\Gamma_\psi^\chi}$.

\step Apply $\cF_{q-1}^\dagger$ to the second register to obtain the
state $\ket{\chi}_*\ket{\Kl_\psi^\chi}.$

\endalgorithm

\goodbreak

\section Computing single Kloosterman sums

In this section, we prove Theorem~\ref{theorem:single}.  As before,
let $\F$ be finite field, let $\psi$ be a non-trivial additive
character of~$\F$, and let $\chi$ be a multiplicative character
of~$\F$.

We can summarise Theorem~\ref{theorem:vector} by saying that there
exists a unitary operator $A$ acting on an $n$-qubit register such
that
$$
A\ket{0} = \ket{\Kl_\psi^\chi}.
$$
Consider any $a\in\F^\times$.  By \eqref{eq:Kl}, the quantum amplitude
of the state $\ket{a}_\times$ in $\ket{\Kl_\psi^\chi}$ is
$$
\kappa(a) = {1\over\sqrt{N_\chi}}\Kl_\psi(a,\chi),
$$
and our goal is to estimate $\kappa(a)$.  To do this, we first
initialise an auxiliary qubit $\ket{u}$ to the state
${1\over\sqrt{2}}(\ket{0}+\rho\ket{1})$, where $\rho$ is a complex
number of absolute value~$1$ that will be chosen below.  We initialise
an $n$-qubit register conditionally on the auxiliary qubit~$\ket{u}$
by either setting it to $\ket{a}_\times$ (if $u=0$) or applying the
operator~$A$ (if $u=1$).  This gives the state
$$
{1\over\sqrt{2}}(\ket{0}\ket{a}_\times+\rho\ket{1}\ket{\Kl_\psi^\chi}).
$$
We then apply a Hadamard transform to the auxiliary qubit.  In total,
we obtain a unitary operator $B_{a,\rho}$ acting on a $1$-qubit
register and an $n$-qubit register and satisfying
$$
B_{a,\rho}(\ket{0}\ket{0})=
{1\over2}\left(
\ket{0}\ket{a}_\times+\rho\ket{0}\ket{\Kl_\psi^\chi}+
\ket{1}\ket{a}_\times-\rho\ket{1}\ket{\Kl_\psi^\chi}\right).
$$
The quantum amplitude of $\ket{0}\ket{a}_\times$ in this state equals
$(1+\rho\kappa(a))/2$.

Let real numbers $d,e\in(0,1)$ be given.  Using amplitude
estimation \citex{BHMT}{Theorem~12} with parameters
$$
k=\left\lceil 1+{1\over2d}\right\rceil
\quad\hbox{and}\quad
M=\left\lceil{k\pi\over\sqrt{1+e}-1}\right\rceil,
\eqnumber{eq:kM}
$$
we can compute an approximation to the probability amplitude
$$
P_{a,\rho}=|1+\rho\kappa(a)|^2/4
$$
that has absolute error at most $e$ with probability at least $1-d$.
Here $M$ is the number of evaluations of $B_{a,\rho}$ performed by the
amplitude estimation algorithm; from the choice of~$M$
in~\eqref{eq:kM}, we deduce that this satisfies $M=O(1/(de))$.

We now take $\rho\in\{1,\zeta,\zeta^2\}$, where $\zeta\in\C$ is a
primitive cube root of unity.  A straightforward computation shows
$$
\kappa(a) = {4\over3}
\bigl(P_{a,1}+\zeta^2 P_{a,\zeta} + \zeta P_{a,\zeta^2}\bigr).
$$
We can therefore compute an approximation to~$\kappa(a)$ that has
absolute error at most $4e$ with probability at least $(1-d)^3$ from
approximations of $P_{a,1}$, $P_{a,\zeta}$, $P_{a,\zeta^2}$ as above.
Multiplying by $\sqrt{N_\chi}/\sqrt{q}<\sqrt{q}$, we obtain an
approximation of $\Kl_\psi(a,\chi)/\sqrt q$ that has absolute error
at most $4e\sqrt{q}$ with probability at least $(1-d)^3$.

For any $\delta,\epsilon\in(0,1)$, taking $e=\epsilon/(4\sqrt{q})$ and
$d=1-(1-\delta)^{1/3}$, and taking $M$ as in~\eqref{eq:kM}, we obtain
an approximation of $\Kl_\psi(a,\chi)/\sqrt{q}$ that has absolute
error at most $\epsilon$ with probability at least $1-\delta$, using
$O(\sqrt{q}/(\delta\epsilon))$ applications of the operators
$B_{a,\rho}$ for $\rho\in\{1,\zeta,\zeta^2\}$.  This implies
Theorem~\ref{theorem:single}.

\goodbreak

\unnumberedsection References

\parindent=\normalparindent
\advance\parskip by1ex

\reference{Biasse-Song} J.-F. {\sc Biasse} and F. {\sc Song}, A
polynomial time quantum algorithm for computing class groups and
solving the principal ideal problem in arbitrary degree number fields.
In: {\sl Proceedings of the Twenty-Seventh Annual ACM-SIAM Symposium
on Discrete Algorithms\/}, 893--902.  ACM, New York, 2016.

\reference{BHMT} G. {\sc Brassard}, P. {\sc H{\o}yer}, M. {\sc Mosca}
and A. {\sc Tapp}, Quantum amplitude amplification and estimation.
{\sl Quantum computation and information\/} (Washington, DC, 2000),
53--74.  Contemporary Mathematics, 305.  American Mathematical
Society, Providence, RI, 2002.

\reference{Childs} A. M. {\sc Childs}, On the relationship between
continuous- and discrete-time quantum walk.  {\it Communications in
Mathematical Physics\/} {\bf 294} (2010), no.~2, 581--603.

\reference{Childs-Schulman-Vazirani} A. M. {\sc Childs}, L. J. {\sc
Schulman} and U. {\sc Vazirani}, Quantum algorithms for hidden
nonlinear structures.  In: {\sl Proceedings of the 48th Annual IEEE
Symposium on Foundations of Computer Science (FOCS 2007)\/}, 395--404.
IEEE Computer Society, Los Alamitos, CA, 2007.

\reference{EHKS} K. {\sc Eisentr\"ager}, S. {\sc Hallgren}, A. {\sc
Kitaev} and F. {\sc Song}, A quantum algorithm for computing the unit
group of an arbitrary degree number field. In: {\sl STOC'14:
Proceedings of the 2014 ACM Symposium on Theory of Computing\/},
293--302.  ACM, New York, 2014.

\reference{Hallgren} S. {\sc Hallgren}, Fast quantum algorithms for
computing the unit group and class group of a number field. In: {\sl
STOC'05: Proceedings of the 37th Annual ACM Symposium on Theory of
Computing\/}, 468--474.  ACM, New York, 2005.

\reference{Hoyer} P. {\sc H{\o}yer}, On arbitrary phases in quantum
amplitude amplification.  {\it Physical Review A\/} {\bf 62}, 052304
(2000).

\reference{Katz} N. M. {\sc Katz}, {\sl Gauss sums, Kloosterman sums,
  and monodromy groups\/}.  Annals of Mathematics Studies, 116.
Princeton University Press, Princeton, NJ, 1988.

\reference{Mosca-Zalka} M. {\sc Mosca} and C. {\sc Zalka}, Exact
quantum Fourier transforms and discrete logarithm algorithms.  {\it
International Journal of Quantum Information} {\bf 2} {2004}, no.~1,
91--100.

\reference{Salie} H. {\sc Sali\'e}, \"Uber die Kloostermanschen Summen
$S(u,v;q)$.  {\it Mathematische Zeitschrift\/} {\bf 34} (1932), no.~1,
91--109.

\reference{Shor} P. W. {\sc Shor}, Polynomial-time algorithms for
prime factorization and discrete logarithms on a quantum computer.
{\it SIAM Journal on Computing\/} {\bf 26} (1997), no.~5, 1484--1509.

\reference{Shparlinski} I. E. {\sc Shparlinski}, Open problems on
exponential and character sums.  In: {\sl Number Theory: Dreaming in
Dreams --- Proceedings of the 5th China-Japan Seminar}, 222--242,
Series Number Theory and its Applications, 6.  World Sci. Publ.,
Hackensack, NJ, 2010.

\reference{Shparlinski-web} I. E. {\sc Shparlinski}, Open problems on
exponential and character sums,\hfill\break {\tt
http://web.maths.unsw.edu.au/\~{}igorshparlinski/CharSumProjects.pdf}

\reference{van Dam-Seroussi} W. {\sc van Dam} and G. {\sc Seroussi},
Efficient quantum algorithms for estimating Gauss sums.  Preprint,
{\tt https://arxiv.org/abs/quant-ph/0207131}.

\reference{Weil} A. {\sc Weil}, On some exponential sums.  {\it Proc.\
Nat.\ Acad.\ Sci.\ U. S. A.\/} {\bf 34} (1948), 204--207.

\vfill

\leftline{Peter Bruin}
\leftline{Universiteit Leiden}
\leftline{Mathematisch Instituut}
\leftline{Postbus 9512}
\leftline{2300 RA \ Leiden}
\leftline{Netherlands}
\leftline{\tt P.J.Bruin@math.leidenuniv.nl}

\bye